\documentstyle[11pt,epsfig,twocolumn]{article}
\newcommand{\beq}{\begin{equation}}
\newcommand{\eeq}{\end{equation}}
\newcommand{\beqa}{\begin{eqnarray}}
\newcommand{\eeqa}{\end{eqnarray}}
\newcommand{\ba}{\begin{array}}
\newcommand{\ea}{\end{array}}
\newcommand{\half}{\frac{1}{2}}

\newtheorem{teo}{Theorem}

\newcommand{\be}{\begin{equation}}
\newcommand{\ch}{\choose}
\newcommand{\ee}{\end{equation}}
\newcommand{\bt}{\begin{teo}}
\newcommand{\et}{\end{teo}}
\newcommand{\s}{\sigma}
\newcommand{\sq}{\sqrt{2 E r^2 + 2 \alpha r - L^2}}

\begin{document}
%   Headings
    \markright{\rm \hfill V.Romanovski,  M. Robnik:
    On WKB Seies for the Radial Kepler Problem  \hfill}
    \pagestyle{myheadings}
    \thispagestyle{plain}
%   End of headings
\twocolumn[\hsize\textwidth\columnwidth\hsize\csname %
@twocolumnfalse\endcsname
% CAMTP preprint comment
  \vspace*{-1.8cm}
  \begin{center}
  \epsfig{file=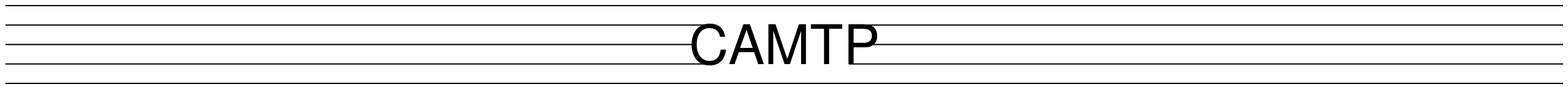,height=9mm,width=\textwidth}\vspace{2mm}

  \noindent Submited to {\it Nonlinear Phenomena in Complex Systems}
  \hfill
   Preprint CAMTP/99-2
   \end{center}\vspace{6mm}
% END of comment
\begin{center}
{\LARGE \bf On WKB Series for the Radial Kepler Problem}
\\
\vspace{0.4cm}

{\large
Valery Romanovski and
Marko Robnik}
\vspace{0.15cm}

{\small \it  Center for Applied Mathematics and Theoretical Physics,
 University of Maribor,\\
 Krekova 2, SI-2000 Maribor, Slovenia\\
 E-mail: robnik@uni-mb.si  and
         valery.romanovsky@uni-mb.si \\
{\rm\small (Submited 12  August  1999)}
}
\end{center}
%\title{ON WKB SERIES FOR THE RADIAL KEPLER PROBLEM }
%\author{ Valery Romanovsky  and  Marko Robnik\\
% Center for Applied Mathematics and Theoretical Physics,\\
%University of Maribor, Krekova 2, SI-2000 Maribor, Slovenia\\
%E-mail: robnik@uni-mb.si}
%\date{}
%\maketitle

\parindent 0pt
\parskip 4pt

\begin{quotation}{
 We obtain the rigorous  WKB expansion
 to all orders for the  radial Kepler problem,
using the residue calculus in evaluating the WKB quantization condition
in terms of a complex contour integral in the complexified coordinate
plane.
The procedure yields the exact energy spectrum of this Schr\"odinger
eigenvalue problem and thus resolves the controversies around the
so-called "Langer correction".
The problem is nontrivial also because there are
only a few systems for which all orders of the WKB series can be
calculated, yielding a convergent series whose sum is equal to the
exact result, and thus sheds new light to similar and more difficult
problems.
\parskip 10pt
\parindent 0pt

\vspace{0.5cm}
}
\end{quotation}
]

In recent years many studies have been devoted to systematical
investigation of the accuracy of semiclassical approximations,
which is a very important problem, especially in the  context of quantum

chaos (Casati and Chirikov  1995, Gutzwiller  1990, Robnik 1998).
Among them  the WKB expansions to all orders  for
a one dimensional  system with the potential $V(x)=U_0/\cos^2 (\alpha
x)$
was investigated by Robnik and Salasnich (1997a) and for the angular
momentum operator by Robnik and Salasnich (1997b) and by
 Salasnich and Sattin (1997). An important earlier classic paper is by
Bender  {\it et al} (1977).
It has been shown for the potentials treated in the above papers  that
all WKB orders can be obtained, that the
semiclassical
series for the eigenvalues converges and when summed yields  the well
known exact results. In the present paper we apply the same WKB method
to the radial Kepler
problem. We solve the problem in a rigorous way, using the residue
calculus
in evaluating the complex contour integrals in order to obtain all
orders
in the $\hbar$-expansion, which confirms the conjecture by Robnik and
Salasnich (1997b) about the form of all WKB terms in this problem.
When summed, the WKB series yields the exact energy spectrum of the
Kepler problem.

The problem is very important as it resolves the controversies
around the so-called "Langer correction" (Langer 1937,
Gutzwiller 1990), which states that in applying the first (leading)
order
WKB method (the torus quantisation of Einstein
and Maslov) one should replace the exact value of the
angular momentum $L=l(l+1)\hbar$ by $L=(l+\half)^2\hbar$,
in which case we then get (by brute force) the exact result for
the Coulomb energy spectrum. Of course, as explained by
Robnik and Salasnich (1997b), this "Langer correction" has no
physical foundation and indeed is just an {\em ad hoc} guess.
Indeed, they have shown that higher order terms in the
Kepler (Coulomb) energy eigenvalue problem exactly compensate
the error done by the "Langer correction", so that after
summing up the infinite WKB series we get the exact result free of
any ad hoc assumptions like "Langer correction". Thus, the
importance of the present paper is to provide a rigorous
proof of the hypothesis (ansatz) of Robnik and Salasnich (1997b)
on this problem. We shall consider the Schr\"odinger
equation in the complexified coordinate space and shall
use the standard WKB ansatz. Then we solve the quantisation
condition (which is equivalent to the single-valuedness
condition of the wavefunction in the complexified coordinate
plane) by using the residue calculus. The problem is not
trivial, because it is hard to guess the correct form
of the polynomial terms that enter the quantisation condition
at each order, and experience in  problems
connected to the polynomial differential equations (Dolichanin
{\em et al} 1998) proved to be crucial.
In fact, using this experience we went
further producing the more general results on WKB method
in the Hamiltonian systems with one degree of freedom
which will be presented in a separate paper (Robnik
and Romanovski 1999).

We consider the Schr\"odinger equation for the radial Kepler  problem
\be   \label{sch}
 [- \frac{\hbar ^2}2 \frac {{\rm d}^2}{{\rm d} r^2}+V(r)]\psi
(r)=E\psi(r)
\ee
where
\be \label{pot}
V(r)=\frac{L^2}{2r^2}-\frac \alpha r
\ee
with $L^2=l(l+1)\hbar^2, l=0,1,2, \dots,$ being the squared angular
momentum,
   and $ \alpha>0.$

We can always write the wavefunction as
\be
\psi(r) =\exp \left\{\frac {\rm  i}\hbar \sigma (r)\right\}
\ee
where the phase $\sigma (r)$ is a complex function that satisfies the
differential equation
\be \label{w1}
\sigma '^2(r)+\left(\frac \hbar {\rm  i} \right) \sigma '' (r)=
2(E-V(r)).
\ee
The WKB expansion for the phase is
\be \label{w2}
\sigma(r)=\sum_{k=0}^{\infty}
\left(\frac \hbar {\rm  i} \right)^k \sigma_k (r).
\ee
Substituting (\ref{w2}) into (\ref{w1}) and comparing
like powers of  $\hbar$ gives the recursion relation
\be \label{w3}
\begin{array}{c}
\s _0'^2 =2 (E-V(r)),\\ \sum_{k=0}^n \s' _k \s ' _{n-k}+\s
_{n-1}''=0.
\end{array}
\ee

The quantization condition is obtained by requiring the uniqueness of
the wavefunction, so for the contour integral around a closed contour in

the complexified
coordinate plane $r$ we must have
\be \label{ci}
\oint_\gamma {\rm d} \sigma  =\sum_{k=0}^{\infty}
 \left(\frac \hbar{\rm  i}
 \right)^k \oint _\gamma{\rm d} \sigma_k = 2\pi n_r \hbar,
\ee
where $n_r\ge 0$, an integer number, is the radial quantum number
and $\gamma $ is a complex  contour enclosing  the turning points on the

real axis.

The zero-order term is given by
\begin{eqnarray} \nonumber
\lefteqn{\oint _\gamma {\rm d} \sigma_0 =2 \int {\rm d} r \sqrt{2
(E-V(r))}}\\
& & =  \alpha
 \pi \sqrt{\frac 2{-E}} -2 \pi L
\end{eqnarray}
and the first odd term is
\be
 \left(\frac \hbar{\rm  i}
 \right) \oint _\gamma{\rm d } \sigma_1 =-\pi \hbar.
\ee

>From (\ref{w3}) we get
\be \label{sig2}
\begin{array}{c}
\s'_2=
  ( L^4 - 6 \alpha  L^2 r
    + (3
    \alpha^2 -12 L^2 E)  r^2 +\\  8 \alpha E r^3)/({
  8 r (\sq)^5 })
\end{array}
\ee
and
\be \label{sig3}
\begin{array}{c}
\s'_3=(3 r [\alpha^3 r +
      2 \alpha E r
       (-5 L^2 + 4 E r^2) +\\
      \alpha^2 (-3 L^2 +
         4 E r^2) -
      2 E L^2
       (L^2 +\\  8 E r^2)])/(8 (\sq)^8).
\end{array}
\ee
Computing also $\s'_4$
one finds,  correspondingly (Robnik and Salasnich 1997b),
\be
 \left(\frac \hbar{\rm  i}
 \right)^2 \oint _\gamma{\rm d } \sigma_2 =- \hbar^2 \frac{\pi}{4L}
\ee
and
\be
 \left(\frac \hbar{\rm  i}
 \right)^4 \oint _\gamma{\rm d } \sigma_4 =\hbar^4 \frac{\pi}{64L^3}.
\ee
>From this result the following general formula for the $\hbar^{2k}$ term

 was conjectured (Robnik and Salasnich 1997b), $k=1,2,3,\dots$
\be \label{f}
 \left(\frac \hbar{\rm  i}
 \right)^{2k} \oint _\gamma{\rm d } \sigma_{2k} =-2 \pi \hbar {\frac 12
\ch k} 2^{-2k} \lambda^{1-2k}
\ee
and so, taking into account that due to the conjecture of Bender  {\it
at al} (1977)
\be \label{15}
\oint_\gamma {\rm d} \sigma_{2k+1}=0,
\ee
we get
\be \label{ff}
\frac \alpha {\sqrt{-2E}}=\hbar [(n_r+\frac 12)+\sum_{k=0}^\infty
{\frac 12 \ch k} 2^{-2k} \lambda^{1-2k}],
\ee
where $\lambda =L/\hbar=\sqrt{l(l+1)},$
implying the energy spectrum for the quantal Kepler problem
(\ref{sch}), (\ref{pot}),
\be \label{ener}
E=\frac{-\alpha^2}{2 \hbar^2(n_r+l+1)^2}.
\ee

We will show that formulae (\ref{f}) and (\ref{15}) indeed hold.
To compute the contour integrals (\ref{ci})
\be \label{int}
 \oint _\gamma{\rm d } \sigma_k
\ee
we  find the form of the  functions $\s' _k$. We have from (\ref{w3})
\begin{eqnarray} \nonumber
\lefteqn{\s' _0=\sqrt{\frac{2 E r^2 + 2 \alpha r - L^2}{r^2}}=}\\ & &
\frac{\sq}r.
\end{eqnarray}
We see that $\s' _0$ is a two valued function.
However we can cut the complex plane along the real axis between the
turning
points $r_1, r_2$ ($r_1<r_2$), which are the roots of the equation
\be
2 E r^2 + 2 \alpha r - L^2=0
\ee
and in such case we obtain a single valued function in the cutting
plane.
Then from (\ref{w3}) we get
\be
\s' _1=  \frac{ \alpha r-L^2  }{2 r (\sq)^2}.
\ee

We now make an "educated guess" concerning the form of the
coefficients $\sigma'_k$. Namely, we shall show that the
formulae (22)--(24) are crucial for our analysis, however it is
not obvious from the first few expressions for $\sigma'_k$
that this shape of the coefficients contains  exactly the
information which we need. It is interesting to mention
that similar recurrence relations arise in the investigation
of the famous problem of ordinary differential equations,
the so-called center-focus problem (Dolichanin {\em et al}
1998, eq.(15)),
and our previous experience turned out to be very helpful here.
Our surmise reads then:

\be \label{s2k}
\s' _{2s}={{P_{4s-1}}\over r( \sq)^{6s-1} },
\ee
and
\be \label{s2k1}
\s'_{2s+1}={{r Q_{4s-1}}\over  { (\sq) ^{6s+2}} },
\ee
where $s>0$,  the degrees of polynomials  $P_{4s-1}, Q_{4s-1}$ (as
functions
of $r$)
  are not greater than  $4s-1$,
and $P_{4s-1}$ has  the form
\be \label{p}
P_{4s-1}=\frac 1 {2^{2s}} {\frac 12 \ch s} (L^{4s}-6 L^{4s-2}\alpha s r
+\dots).
\ee
Here and below we denote by $Q_k$ any polynomial of degree less than or
equal to $k$.

We prove the formulae (\ref{s2k})--(\ref{p}) by induction on $s$.
According to (\ref{sig2}), (\ref{sig3}) we see that for $s=1$ the
statements are true.
Let us suppose that they are proven for all $s<m$ and consider the case
$s=m$.

To compute   $\s'_{2m}$
 using the recurrence formula (\ref{w3})  we note that
\begin{eqnarray} \nonumber
\lefteqn{ \frac { \s'_{2k+1} \s'_{2(m-k-1)+1}}{ \s'_0}=}\\& & \frac{r^3
Q_{4k-1}
  Q_{4m-4k-5}}{ (\sq)^{6m-1}} \label{1} \\ & &
= \frac{r^4  Q_{4m-6}}{r  (\sq)^{6m-1}},\nonumber
\end{eqnarray}
\begin{eqnarray}\label{2}
\lefteqn{\frac { \s'_{1} \s'_{2(m-1)+1}}{ \s'_0}=}\\ & &
 \frac{r^2 ( \alpha r- L^2) Q_{4m-5}}{ r (\sq)^{6m-1}} \nonumber
\end{eqnarray}
and
\begin{eqnarray} \label{3}
\lefteqn{\hspace{1.5cm}   \frac {\s_{2m-1}''}{\s'_0}=}\\ & &
\left(\frac{r  Q_{4m-5}}{
(\sq)^{6m-4}}\right)' \frac 1{\s'_0}=\nonumber  \\ & &
( r^2 [   (-L^2 + 6    \alpha r -
 6 \alpha m r +\nonumber  \\ & &
           10 E r^2 - 12 E m r^2) Q_{4m-5}  + \nonumber  \\
& &
  r (-L^2 +   2 \alpha r +  E r^2)\times \nonumber  \\ & &
Q_{4m-5}'])/({r (\sq )^{ 6  m-1 }}). \nonumber
\end{eqnarray}
Thus we see that the fractions (\ref{1})-- (\ref{3}) have the form
(\ref{s2k}) and the polynomials in the numerators have no  influence on
the two lowest terms of the numerator of $\s'_{2m}$.

Therefore
\begin{eqnarray} \label{pp}
\lefteqn{\s'_{2m} =-\sum_{k=1}^{m-1}\frac{ \s'_{2k}
 \s'_{2(m-k)}}{2\s'_0}+}\\& &  \frac{r^2 Q_{4m-3} }{r( \sq)
^{6m-1}}=\nonumber\\ & &
- \sum _{k=1}^{m-1}
\frac{P_{4k-1} P_{4(m-k)-1}+r^2 Q_{4m-3} }{2r(\sq)^{6m-1}}.\nonumber
\end{eqnarray}
Note that the degree of the polynomials $ P_{4k-1} P_{4(m-k)-1}$
is less or  equal $4m-2$.
Hence it  remains to compute the two lowest terms of the polynomial
$\s'_{2m}.$
Observing now that
\begin{eqnarray} \nonumber
\lefteqn{P_{4k-1} P_{4(m-k)-1}=\frac 1{2^{2m}}{\frac 12 \ch k} {\frac 12
\ch
{m-k}} } \\ & & \times (L^{4m}-6m\alpha r L^{4m-2}+\dots )
\end{eqnarray}
we get from (\ref{pp})
\begin{eqnarray}
\lefteqn{\s'_{2m}=}\\& & -
\frac{L^{4m}-6 L^{4m-2}m \alpha r+r^2 Q_{4m-3} }{2^{2m+1}r(\sq)^{6m-1}}
\nonumber \\& & \times
\sum _{k=1}^{m-1}{\frac 12 \ch k}{\frac 12 \ch m-k}. \nonumber
\end{eqnarray}
Therefore, taking into account the equality
\be
\sum _{k=0}^{m} {\frac 12 \ch k}{\frac 12 \ch m-k }=0
\ee
(which is just a special case of the Vandermonde's convolution,
see e.g. (Graham {\it at al} 1994))
we conclude that (\ref{s2k}) holds.

It remains to find   $\s'_{2m+1}$. We get
\begin{eqnarray}
\lefteqn  {-\frac { \s'_{1} \s'_{2m}}{ \s'_0}-
\frac{\s_{2m}''}{2 \s'_0} =} \label{31}\\ \nonumber
& &  (( 6 \alpha m + 12 E m r)P_{4m-1}
 +\hspace{1cm} \nonumber  \\ & &  (L^2 -
  2 \alpha r -   2  E r^2)\times \hspace{2cm} \nonumber  \\
    & &  P_{4m-1}' )/({2( \sq) ^{6m+2}}) \nonumber
\end{eqnarray}
and in the case $0<k<m$
we have
\begin{eqnarray}
\lefteqn{\frac { \s'_{2k} \s'_{2(m-k)+1}}{ \s'_0}=} \label{32} \\ & &
\frac{r P_{4k-1}
Q_{4m-4k-1}}{ (\sq)^{6m+2}}. \nonumber
\end{eqnarray}

Using the induction assumptions (\ref{s2k}) and (\ref{s2k1})
  we conclude that (\ref{31}) and (\ref{32})
 are of the form (\ref{s2k1}), so each $\s'_{2m+1}$ obeys (\ref{s2k1}).

Now we can  compute the  integrals (\ref{int}). We are going to use
the residue calculus. The integration contour in our complex $r$
plane (complexification of the coordinate space $r$) is chosen so that
it
encircles the two turning points $r_1$ and $r_2$.
They are connected by a cut so that each $\s'_m$ is single-valued,
and has singularities (poles) either at infinity or at $r=0.$
Hence the integration contour encircles now the poles that lie outside
the integration contour and, therefore, the  calculation becomes very
easy. Namely,
using the  residue theorem we get
\begin{eqnarray} \label{ur}
\lefteqn{ \left(\frac \hbar{\rm  i}
 \right)^{2k} \oint _\gamma{\rm d } \sigma_{2k} =}\\& &   2\pi {\rm i}
 \left(\frac \hbar{\rm  i}
 \right)^{2k} ({\rm Res} _0 \s'_{2k}+{\rm Res} _\infty
\s'_{2k}).\nonumber
\end{eqnarray}
>From (\ref{s2k}), (\ref{p})
it is easily seen
\be
{\rm Res}_0 \s'_{2k}= \frac{{\rm i}}{2^{2k} {\rm i}^{6k} L^{2k-1}}
 {\frac 12 \ch k}.
\ee

For large $r$ we have according to  (\ref{s2k}) the convergent  series
\begin{eqnarray}\nonumber  \lefteqn{
\s'_{2k}=\frac{P_{4k-1}}{ r \left( r\sqrt{2E} \sqrt{1 + \frac  \alpha{E
r}
 -\frac{ L^2}{2Er^2}}\right)^{6k-1}}} \\ & & =\frac {P_{4k-1}}{r (r
\sqrt{2E})^{6k-1}}\times \\ & &
[1+(6k-1)(\frac{\alpha}{Er }-\frac{L^2}{2 Er^2})+\dots].\nonumber
\end{eqnarray}

and, therefore,

\be
{\rm Res}_\infty \s'_{2k}=0.
\ee
Similarly, we obtain
\be
 \oint _\gamma{\rm d } \sigma_{2k +1}=0,
\ee
since $r=0$ is not a singular point of $\s'_{2k+1}$ and
\be
{\rm Res} _\infty \s'_{2k+1}=0.
\ee
Hence
\begin{eqnarray}
\lefteqn{ \left(\frac \hbar{\rm  i}
 \right)^{2k} \oint _\gamma{\rm d } \sigma_{2k} =}\\ & &  2\pi {\rm i}
\frac {\hbar^{2k}}{(-1)^k}
 \left(\frac {{\rm i}}{(-1)^k 2^{2k} L^{2k-1}} {\frac 12 \ch k}
 \right)=\nonumber \\& & -2\pi \hbar \frac {\lambda ^{1-2k}}{2^{2k}}
{\frac 12 \ch k}. \nonumber
\end{eqnarray}
Thus formulae (\ref{f}), (\ref{15}) and (\ref{ff}) are proven, resulting

in the exact
energy spectrum (\ref{ener}).

In conclusion, we have shown that the residue theorem
can be successfully applied in computing the WKB series of certain
potentials  in one degree of freedom, in particular for the
Kepler (Coulomb) potential. In this way we have confirmed the
conjecture by Robnik and Salasnich (1997b) about the form of
all the terms in the corresponding WKB expansion and thus
resolved the controversies about the so-called "Langer correction"
(Langer 1937, Gutzwiller 1990), by explaining that by
ignoring the "Langer correction" and assuming the exact value
of the quantal angular momentum we indeed get the exact
result for the energy spectrum after calculating the terms
of all orders and by summing the WKB series (after completing our
 work the preprint by Hainz and Grabert (1999)
with similar result (but  another convergent WKB expansion) was posted).

  It appears interesting to apply this
method  to the  investigation of other potentials like, for example,
the  above mentioned  potential, $V=U_0/\cos^2 (\alpha x)$. Here the
difficulties arise  because the integrated function has infinitely
 many singular points.

The most important problem in this context is the general formulation
of the exact WKB quantization condition in terms of contour integrals
for arbitrary one-dimensional potentials.

\section*{Acknowledgments}
This project is supported by the Ministry of Science and Technology of
the Republic of Slovenia and by the  Rector's Fund of the University of
Maribor.
VR acknowledges the support of the work by the grant of the Ministry of
Science
and Technology of the Republic of Slovenia and  the Abdus Salam   ICTP
(Trieste) Joint Programme.


\begin{thebibliography}{99}
\bibitem{}
Bender C. M., Olaussen K. and Wang P. S. (1977) {\em Phys. Rev. D} {\bf
16} 1740
\bibitem{}
 Casati G. and Chirikov B. V. (1995) {\it Quantum Chaos}
(Cambridge: Cambridge University Press)
\bibitem{}
Dolichanin Ch.,\ \  Romanovski V.\   and   Stefanovich M. (1998) {\it
Dif\-feren\-tial'nye
Urav\-neniya} (Minsk) {\bf 34} 1587 (in   Russian,  to be translated in
{\it Differential Equations})
\bibitem{}
Graham R. L., Knuth D. E. and Patashnik O. (1994)  {\em Concrete
Mathematics}
(New York: Addison-Wesley Publishing)
\bibitem{}
Gutzwiller M. C. (1990) {\em Chaos in Classical and Quantum Mechanics}
(Berlin:
Springer)
\bibitem{}
Hainz J. and Grabert H. (1990) Centrifugal terms in the WKB
approximation and semiclassical quantization
of hydrogen, preprint, quant-ph/9904103
\bibitem{}
Langer R. I. (1937) {\em Phys. Rev.} {\bf 51} 669
\bibitem{}
 Maslov V. P. and Fedoriuk M. V. (1981) {\it Semi-Classical
Approximation
in Quantum Mechanics} (Dordrecht: Reidel)
\bibitem{}
Robnik M. (1998) {\em Nonlinear Phenomena in  Complex Systems} (Minsk)
{\bf 1} 1
\bibitem{}
Robnik M. and Salasnich L. (1997a) {\em J.Phys. A: Math Gen} {\bf 30}
1711
\bibitem{}
 Robnik M. and  Salasnich L.
 (1997b)  {\em J.Phys. A: Math. Gen.} {\bf 30} 1719
\bibitem{}
 Salasnich L. and Sattin F. (1997) {\em J.Phys. A: Math. Gen.} {\bf 30}
7597
\end{thebibliography}
\end{document}